\RequirePackage{fix-cm}
\documentclass[twocolumn]{svjour3}          
\smartqed  
\usepackage{graphicx}
\usepackage{natbib}

\begin{document}

\title{Prospects for the characterisation of exo-zodiacal dust with the VLTI}


\author{S. Ertel         \and
        O. Absil         \and
        D. Defr\`ere     \and
        J.-C. Augereau   \and
        B. Mennesson
}

\institute{S. Ertel \at
              Steward Observatory, Department of Astronomy, University of Arizona, 993 N. Cherry Ave, Tucson, AZ 85721, USA\\
              \email{sertel@email.arizona.edu}
           \and
           O. Absil \at
              Space sciences, Technologies and Astrophysics Research (STAR) Institute, Universit\'e de Li\`ege, 19c All\'e du Six Ao\^ut, 4000 Li\`ege, Belgium
           \and
           D. Defr\`ere \at
              Space sciences, Technologies and Astrophysics Research (STAR) Institute, Universit\'e de Li\`ege, 19c All\'e du Six Ao\^ut, 4000 Li\`ege, Belgium
           \and
           J.-C. Augereau \at
              Univ. Grenoble Alpes, IPAG, 38000 Grenoble, France\\
              CNRS, IPAG, 38000 Grenoble, France
           \and
           B. Mennesson \at
              Jet Propulsion Laboratory, California Institute of Technology, 4800 Oak Grove Drive, Pasadena CA 91109-8099, USA
}

\date{Received: date / Accepted: date}

\maketitle

\begin{abstract}
Exo-zodiacal dust, exozodi for short, is warm ($\sim$300K) or hot (up to $\sim$2000K) dust
found in the inner regions of planetary systems around main sequence stars. In analogy to
our own zodiacal dust, it may be located in or near the habitable zone or closer in, down to
the dust sublimation distance. The study of the properties, distribution, and evolution of
exozodis can inform about the architecture and dynamics of the innermost regions of planetary
systems, close to their habitable zones. On the other hand, the presence of large amounts of
exo-zodiacal dust may be an obstacle for future space missions aiming to image Earth-like
exoplanets. The dust can be the most luminous component of extrasolar planetary systems,
but predominantly emits in the near- to mid-infrared where it is outshone by the host star.
Interferometry provides a unique method of separating the dusty from the stellar emission. We
discuss the prospects of exozodi observations with the next generation VLTI instruments and
summarize critical instrument specifications.
\end{abstract}

\begin{figure*}
  \centering
  \includegraphics[width=1.0\linewidth]{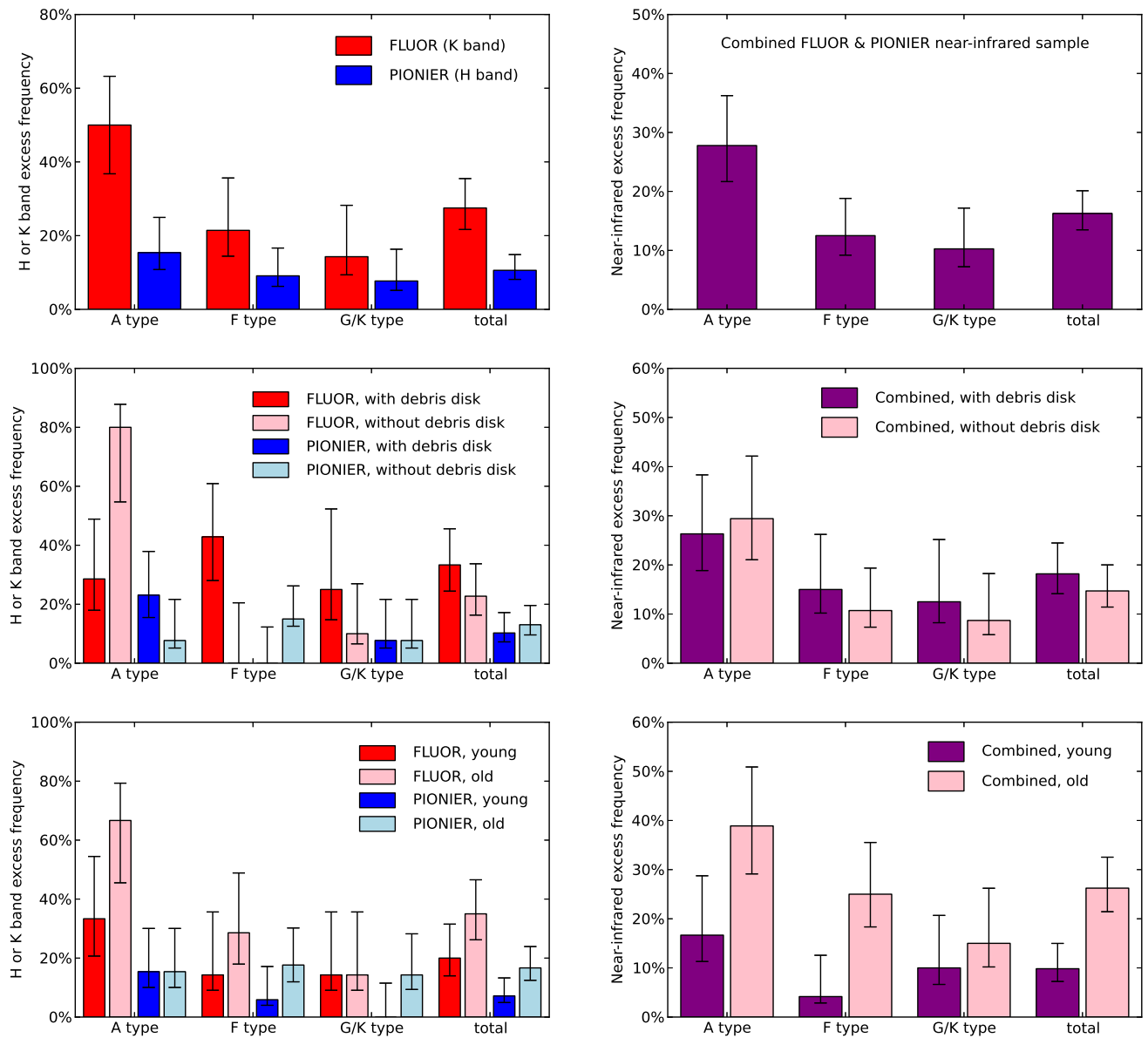}
  \caption{Single (left) and combined (right) statistics from the FLUOR (red) and PIONIER (blue) sample.
The top row shows the excess fraction with respect to the stellar spectral type, the middle
row shows the same, but, in addition, separated for stars with and without a debris disk detected,
and the bottom row shows the excess fraction for different spectral types and separated for stars
younger and older than the median age in each spectral type bin. For details and interpretation
see \citet{ert14b}.}
  \label{fig_stats}
\end{figure*}

\section{Introduction}
The zodiacal dust in our Solar system is distributed between a few AU from the Sun down to
$\sim$4~Solar radii (where it forms the F-corona) with a shallow surface density profile
\citep{kim98, hah02}. In analogy to this dust, we call any dust near the habitable zone
of other stars, or closer in, exo-zodiacal dust (or exozodi for short). In other systems
we usually observe two distinct types of dust. Warm dust near the habitable zone is detected
in the mid-infrared \citep{mil11, men14, ert18}, while hot dust very close to the star
is detected in the near-infrared \citep{abs06, abs13, ert14b}. The connection between the
two is not clear and the distinction might be largely observational. However, no correlation
between the detections at the two wavelengths has been found so far,
which would be expected from simply scaling the Solar system's dust distribution to the
higher dust levels detectable in other systems. This suggests that other systems have
diverse architectures and that the two species might or might not be directly connected.

The spatial dust distribution relates to its location of origin, production mechanism (asteroid
collision, comet evaporation), and the dynamical interaction with the stellar radiation, wind,
and magnetic field, as
well as with potential planets. It thus traces the architecture and dynamics of the inner
regions of planetary systems, in particular that of asteroids and comets \citep{bon12, bon13b,
bon14, leb13, far17} that are otherwise hard or impossible to study. On the other hand, the
presence of large amounts of dust in or near the habitable zone of planetary systems may
be an obstacle for the imaging detection and characterization of Earth-like, potentially
habitable exoplanets (exo-Earths) by future space missions. The dust emission causes
additional photon noise in the observations and clumps in the dust distribution, albeit
potentially caused by interaction with planets, can be misinterpreted as actual planets
given the limited angular resolution and signal-to-noise ratio expected from such
observations \citep{def10, rob12, sta15}. Even hot dust close to the star may significantly
degrade coronagraphic performance at the level needed for exo-Earth imaging, as it produces
emission that is more extended than the star and consequently cannot be perfectly suppressed.
The study of the incidence rate of exo-zodiacal dust, its correlation with other, more
accessible properties of the systems such as stellar luminosity, age, and the presence of
an outer, cold reservoir of dust (debris disk, e.g. \citealt{eir13}), and of the dust
properties is critical for the success of such missions.

While rather small amounts of this dust can be relatively luminous due to its large surface
and high temperature of few 100 to $\sim$2000\,K, the emission of such warm or hot dust peaks
in the near- to mid-infrared (nIR to mIR), where the emission of the host star dominates. The
emission from the exozodis found so far produces an excess at the 1\% level and below. Thus,
detecting the dust requires spatially resolving the dust in order to disentangle its emission
from that of the star. Because of the small angular scales (1\,AU at a distance of 10\,pc
corresponds to an angular separation from the host star of 0.1\,arcsec), only interferometry is
currently able to reach the required angular resolution. Near-IR interferometry has  produced
the largest number of detected exozodis to date. The VLTI instrument PIONIER has been particularly
efficient in surveying a large number of stars. The broad wavelength coverage of the
second-generation suite of VLTI instruments (and PIONIER) is particularly well suited for the
characterization of exozodis.
In this article, we discuss the prospects of observing exo-zodiacal dust with these instruments.
We briefly summarize the past and current observational efforts to study exozodis with focus
on interferometric observations (Sect.~\ref{sect_current}).
For a detailed review of the field of exo-zodiacal dust, we refer
to \citet{kra17}. We explain the method used to detect exozodis with optical long baseline
interferometry (Sect.~\ref{sect_method}) which is the relevant method for all current and near
future VLTI instruments. In Sect.~\ref{sect_prospects}, we discuss the prospects of using
the second generation VLTI instruments to further study exo-zodiacal dust. The critical
technical requirements of instruments for exozodi observations are reviewed in
Sect.~\ref{sect_specs}. We present our conclusions in Sect.~\ref{sect_conclusions}.

\section{Recent observational success}
\label{sect_current}
The first detection of faint nIR excess at the 1\% level around a main sequence star was
reported for Vega by \citet{abs06}. It was achieved using the FLUOR beam combiner on the
CHARA array. Motivated by this detection, survey observations were carried out using
CHARA/FLUOR and the PIONIER beam combiner on the VLTI that produced the first statistical
samples of detected exozodis (\citealt{abs13, ert14b}; Fig.~\ref{fig_stats}). As of today, more than
200~stars have systematically been surveyed in the nIR with a detection rate of $\sim$20\%
(\citealt{nun17}; Marion et al, subm.). Follow-up observations of a sample of
detections have demonstrated the repeatability of the observations and revealed first signs
of significant variability of the excesses (\citealt{nun17}; \citealt{ert16}). These
observations provide detailed information about the presence of exozodis and their correlation
with various properties of the host systems, and motivated and informed several theoretical
studies (\citealt{bon12, bon13b, bon14, leb13, rie16, far17}; McKinley et al, in prep.).
In parallel, first excesses from warm dust near the habitable zone of other stars were
detected using mIR nulling interferometry on the MMT \citep{sto10} and the Keck Interferometric
Nuller \citep{mil11, men14}. Combining data from both nIR and mIR interferometric observations
provide particularly strong constraints of the dust properties and location \citep{def11, leb13,
kir17}. Recently, the Large Binocular Telescope Interferometer (LBTI) has surveyed a
sample of $\sim$40 nearby main sequence stars for habitable zone dust at unprecedented sensitivity
(\citealt{ert18, ert18b}; Ertel et al, in prep.) and achieved a detection rate similar to those
in the nIR. Strong constraints on the scattered light content
of hot exozodi systems have been provided by parts per million accuracy polarimetric observations
\citep{mar16}. Additional constraints have been provided by ground and space based spectroscopic
observations (e.g., \citealt{che06, lis07, bal14}; Lisse et al, in prep.) and by studying WISE data
of a large sample of stars to constrain the incidence rate of exozodis at the bright end of their
luminosity function \citep{ken13}.

\begin{figure}
  \centering
  \includegraphics[width=\linewidth]{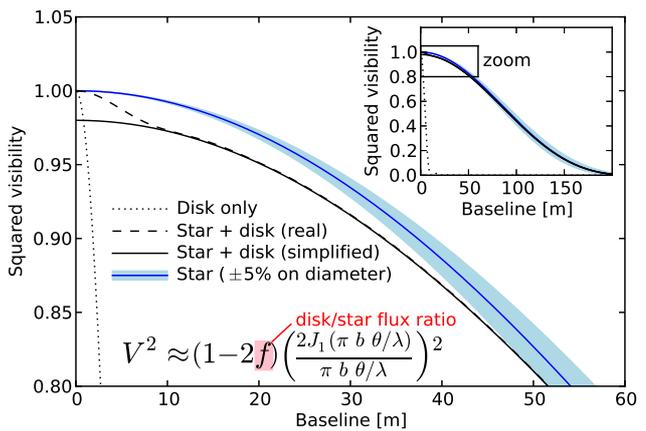}
\caption{Detection strategy of hot exo-zodiacal dust using near-infrared optical interferometry.}
\label{fig_strategy}
\end{figure}

\section{Detection using optical long baseline interferometry}
\label{sect_method}
By far the largest number of exozodis has been detected using optical long baseline interferometry
in the near-infrared as it is used on the VLTI. Using this technique at short baselines (few 10m),
the star remains mostly unresolved, resulting in fully coherent emission. In contrast, the extended
emission from a dust disk is ideally fully resolved, resulting in incoherent emission and
thus a visibility deficit compared to the values expected from the star alone. This visibility
deficit allows for detecting the dust and the disk-to-star flux ratio can be measured as half
the visibility deficit (Fig.~\ref{fig_strategy}). The fact that the star is almost unresolved
is important, because in this case the uncertain stellar diameter has very little impact on the
prediction of the visibilities from the star alone and thus the visibility deficit can be
measured at high accuracy.

Due to the small disk-to-star flux ratio of typically 1\%, only few instruments reach the
accuracy on the visibility measurements necessary to detect the dust. CHARA/FLUOR and
VLTI/PIONIER have been used to survey a large sample of stars for hot exozodis.
PIONIER in combination with the VLTI architecture was proven to be particularly well suited
due to the high efficiency (simultaneous use of four telescopes) and the availability of
closure phase measurements which allow one to directly distinguish between a faint companion
and an extended dust disk as the cause for the detected excess \citep{mar14}.

\section{New opportunities with VLT and LBT interferometry}
\label{sect_prospects}
So far, most detections were achieved only in one band. The large spectral separation
of the $H$/$K$~and the $N$~bands and different inner working angles and sensitivities of the
instruments used make it difficult to connect detections in the two bands or trace the excess
from one band to the other. PIONIER observations have provided a small spectral dispersion
across the $H$~band, but the spectral separation of the channels is too small to put strong
constraints on the spectral shape of the excesses. Only the characterization of exozodis can
answer the three most urgent key questions beyond the frequency and abundance of massive dust
systems:

\begin{itemize}
 \item What is the connection between the warm and hot dust? This is important because most
systems so far have been detected in the nIR, but the implications of this for the presence
of habitable zone dust are unclear. A tentative anti-correlation between the presence of hot and
warm dust has been suggested \citep{men14}, but contradictory results were found by Marion et al
(subm.) based on a different strategy and a larger sample.
 \item Where is the dust located? Detailed temperature information can only be derived from
multi-wavelength observations that sample well the relevant parts of the dust's spectral energy
distribution (nIR to mIR). In addition, spatial information may be derived from observations
at suitable wavelengths and baselines \citep{kir18}. The dust location gives critical insight
into the architecture and dynamics of the underlying planetary system. Furthermore, the dust
location relative to the habitable zone is critical information for future exo-Earth imaging
missions.
 \item What are the dust properties? The dust is detected as thermal emission in the mIR and as
a potential combination of thermal emission and scattered light measured in the nIR. Only with
a detailed knowledge of the dust properties is it possible to estimate from these observations
the amount of scattered light expected in the visible where future exo-Earth imaging missions will
operate.
\end{itemize}

\begin{figure}
  \centering
  \includegraphics[width=\linewidth]{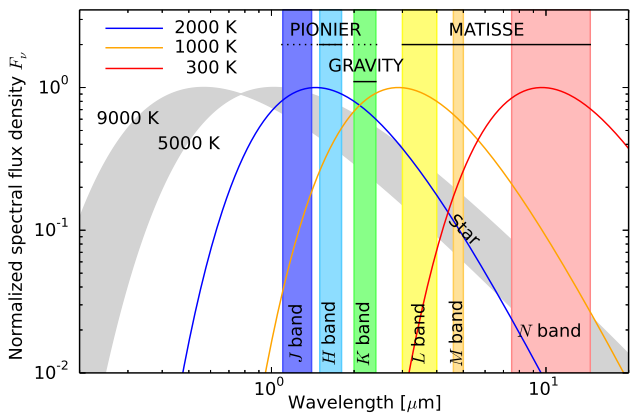}
\caption{Wavelength coverage of the second-generation VLTI instruments (and PIONIER) compared to the
wavelength range in which the emission from hot and warm exo-zodiacal dust is expected to peaks.
The dust emission is approximated by a modified blackbody (a blackbody multiplied by $\lambda^{-2}$), which
represents very small grains as suggested at least for hot dust by PIONIER observations at multiple
wavelengths \citep{ert14b}. For PIONIER,
the dotted range indicates the $K$~band which is no longer available and the $J$~band which may be reached
with a potential instrument upgrade. Towards the shortest bands ($H$~band and in particular $J$~band), scattering
may have a significant contribution to the total emission. The characterization of scattered light is
particularly critical for the understanding of the impact of the dust on exo-Earth imaging missions.}
\label{fig_vlti}
\end{figure}

The second-generation VLTI instruments GRAVITY and MATISSE, together with PIONIER, can provide
the ideal tools to address these questions through multi-wavelength measurements of the spectral
energy distribution (SED) of the excess emission in the nIR to mIR wavelength range
(Fig.~\ref{fig_vlti}). In this range the emission of warm and hot dust peaks and carries most
information about the dust temperature and properties. The well-established detection strategy
used for FLUOR and PIONIER can be employed with all instruments. The broad spectral capabilities
of GRAVITY and MATISSE will allow for strong constraints on the dust properties through a better
characterization of the SED shape and the potential detection of dust emission features (e.g.,
3\,$\mu$m and 10\,$\mu$m silicate features). First steps toward a spectral characterization of the excesses
have been taken with PIONIER \citep{def12, ert14b}. Furthermore, a new survey in a wavelength range
where the dust emits more strongly than in the $H$~band reached with PIONIER will result in a larger
sample to characterize and stronger statistical constraints on the incidence, properties, and
evolution of the dust. It has been shown already, that the detection rate of hot exozodis is about
twice as high in $K$~band compared to $H$~band at similar accuracy of the visibility measurements
\citep{ert14b}. The $K$~band beam combiner GRAVITY and the short wavelength channels of MATISSE
($L$, $M$~bands) will be ideally suited for such surveys assuming a similar accuracy on the measurements
as with PIONIER can be reached on a statistically significant number of stars.

First observations with GRAVITY during science verification and open time have demonstrated the
required accuracy to detect exozodis and demonstrated the wealth of information contained in
the low resolution spectra produced in the K~band (Defr\`ere et al, in prep.).

\section{Critical technical requirements}
\label{sect_specs}
There are three main challenges to be overcome for being able to fully characterize hot and warm
exozodis with the VLTI:

\begin{itemize}
 \item It is critical to reach an accuracy of at least 1\% on the individual measurements of the calibrated
squared visibility. This is necessary to reach a sufficient cumulative accuracy on the source visibilities to
detect the excess detected in available observations and accurately measure the disk-to-star flux ratio.
Previous detections at the 1\% level have been achieved in both the near-infrared and the mid-infrared
(see Sect.~\ref{sect_current}). While the emission from blackbody grains emitting in the near-infrared
at temperatures $<$2000\,K is expected to increase toward longer wavelengths, this is not seen in spectrally
dispersed PIONIER observations (although the constraints on the slope are weak due to the small wavelength
coverage). The presence of very small grains would explain this (Fig.~\ref{fig_vlti}). As a consequence,
the emission is not expected to be above a few percent at any wavelength. While this sensitivity is readily
reached with PIONIER and has been demonstrated for GRAVITY, it is not expected to be reached with MATISSE
without the use of a fringe tracker. MATISSE is thus currently limited to observing very extreme
systems with massive warm dust detections (e.g., \citealt{lis07, lis08, olo12}).
 \item Pointing (field rotation) dependent polarization effects in the VLTI optical path limit the
absolute calibration of single PIONIER observations to ~3\%. A specific observing strategy and additional
calibration of this effect has to be employed to circumvent this limit \citep{ert14b}. This,
however, requires observations of a large number of targets in a consistent way throughout a whole
night which can only be carried out in visitor mode and puts significant limits on the flexibility
of such observations. Correcting this effect on the instrument side (as done with GRAVITY) or
solving it on the VLTI side is critical for efficient and flexible high accuracy observations.
 \item To reach a sufficient cumulative accuracy on the measured source visibilities requires several
measurements of one target. PIONIER has proven to be very efficient due to the simultaneous use of 4
telescopes (6 baselines). Still, several consecutive, calibrated measurements per target are necessary.
Moreover, a potential variability of the hot emission has been detected on a timescale at least as short
as one year \citep{ert16}. The shortest timescale of these variations is not known. This variability
calls at least for quasi-simultaneous observations of a target with all three instruments in order to
avoid corruption of the obtained SED data by excess variability between the different observations.
To significantly increase the efficiency of the observations for both a survey for new systems and
the characterization of known systems, a fully simultaneous use of all three instruments (such as the
``i-SHOOTER'' concept, to use GRAVITY, MATISSE, and PIONIER simultaneously to obtain $H$~to $N$~band
data of a target at the same time) will be highly beneficial.
\end{itemize}

In addition to these critical requirements, further increasing the sensitivity to circumstellar
excess would be highly beneficial for exozodi science with the VLTI. In particular, it would
allow for surveys for exo-zodiacal dust in the Southern hemisphere with a sensitivity similar to
that reached in the North with the LBTI and to detect systems only a few times brighter than our
own zodiacal dust. In the near-infrared, the sensitivity of VLTI observations to faint circumstellar
emission is currently limited by the ability to accurately measure and calibrate visibilities and to
predict the stellar visibilities. This is less critical when aiming to detect dust closer to the habitable
zone of a system in the mid-infrared, because lower angular resolution is sufficient. The star is then even
less resolved and the impact of its uncertain diameter thus further reduced. In addition, nulling interferometry
can be used to remove the stellar light in addition to spatially separating it from the disk emission.
An instrument employing this technique has been introduced as the hi-5 concept by \citet{def18} and can improve the
high contrast capabilities of the VLTI by one order of magnitude. Moreover, such an instrument would be the
first to search for faint habitable zone dust in the Southern hemisphere, critically extending the sample of
stars that can be searched.

\section{Conclusions}
\label{sect_conclusions}

The multi-wavelength capabilities of the second generation VLTI instruments will be vital to further study
exo-zodiacal dust. However, in particular for MATISSE the visibility accuracy needed to detect exozodis
is not expected to be readily met at the time of commissioning. Additional effort will be needed to enable
the exploitation of the VLTI capabilities and to retain its leading role in the characterization of exozodis
in the near future. The gain of one order of magnitude in high-contrast capabilities, as it could be
achieved with a nulling interferometer (hi-5 concept), would allow for a survey for habitable zone dust at
the extreme efficiency demonstrated by PIONIER and at the accuracy currently only reached with lower
efficiency and on a limited sample of stars in the Northern hemisphere.


\bibliographystyle{spbasic}      
\bibliography{bibtex.bib}   

\begin{thebibliography}{36}
\providecommand{\natexlab}[1]{#1}
\providecommand{\url}[1]{{#1}}
\providecommand{\urlprefix}{URL }
\expandafter\ifx\csname urlstyle\endcsname\relax
  \providecommand{\doi}[1]{DOI~\discretionary{}{}{}#1}\else
  \providecommand{\doi}{DOI~\discretionary{}{}{}\begingroup
  \urlstyle{rm}\Url}\fi
\providecommand{\eprint}[2][]{\url{#2}}

\bibitem[{{Absil} et~al(2006){Absil}, {di Folco}, {M{\'e}rand}, {Augereau},
  {Coud{\'e} du Foresto}, {Aufdenberg}, {Kervella}, {Ridgway}, {Berger}, {ten
  Brummelaar}, {Sturmann}, {Sturmann}, {Turner}, and {McAlister}}]{abs06}
{Absil} O, {di Folco} E, {M{\'e}rand} A, {Augereau} JC, {Coud{\'e} du Foresto}
  V, {Aufdenberg} JP, {Kervella} P, {Ridgway} ST, {Berger} DH, {ten Brummelaar}
  TA, {Sturmann} J, {Sturmann} L, {Turner} NH, {McAlister} HA (2006)
  {Circumstellar material in the <ASTROBJ>Vega</ASTROBJ> inner system revealed
  by CHARA/FLUOR}. \aap 452:237--244, \doi{10.1051/0004-6361:20054522},
  \eprint{arXiv:astro-ph/0604260}

\bibitem[{{Absil} et~al(2013){Absil}, {Defr{\`e}re}, {Coud{\'e} du Foresto},
  {Di Folco}, {M{\'e}rand}, {Augereau}, {Ertel}, {Hanot}, {Kervella},
  {Mollier}, {Scott}, {Che}, {Monnier}, {Thureau}, {Tuthill}, {ten Brummelaar},
  {McAlister}, {Sturmann}, {Sturmann}, and {Turner}}]{abs13}
{Absil} O, {Defr{\`e}re} D, {Coud{\'e} du Foresto} V, {Di Folco} E,
  {M{\'e}rand} A, {Augereau} JC, {Ertel} S, {Hanot} C, {Kervella} P, {Mollier}
  B, {Scott} N, {Che} X, {Monnier} JD, {Thureau} N, {Tuthill} PG, {ten
  Brummelaar} TA, {McAlister} HA, {Sturmann} J, {Sturmann} L, {Turner} N (2013)
  {A near-infrared interferometric survey of debris-disc stars. III. First
  statistics based on 42 stars observed with CHARA/FLUOR}. \aap 555:A104,
  \doi{10.1051/0004-6361/201321673}, \eprint{1307.2488}

\bibitem[{{Ballering} et~al(2014){Ballering}, {Rieke}, and
  {G{\'a}sp{\'a}r}}]{bal14}
{Ballering} NP, {Rieke} GH, {G{\'a}sp{\'a}r} A (2014) {Probing the Terrestrial
  Regions of Planetary Systems: Warm Debris Disks with Emission Features}. \apj
  793:57, \doi{10.1088/0004-637X/793/1/57}, \eprint{1407.7547}

\bibitem[{{Bonsor} et~al(2012){Bonsor}, {Augereau}, and {Th{\'e}bault}}]{bon12}
{Bonsor} A, {Augereau} JC, {Th{\'e}bault} P (2012) {Scattering of small bodies
  by planets: a potential origin for exozodiacal dust?} \aap 548:A104,
  \doi{10.1051/0004-6361/201220005}, \eprint{1209.6033}

\bibitem[{{Bonsor} et~al(2013){Bonsor}, {Raymond}, and {Augereau}}]{bon13b}
{Bonsor} A, {Raymond} SN, {Augereau} JC (2013) {The short-lived production of
  exozodiacal dust in the aftermath of a dynamical instability in planetary
  systems}. \mnras 433:2938--2945, \doi{10.1093/mnras/stt933},
  \eprint{1306.0592}

\bibitem[{{Bonsor} et~al(2014){Bonsor}, {Raymond}, {Augereau}, and
  {Ormel}}]{bon14}
{Bonsor} A, {Raymond} SN, {Augereau} JC, {Ormel} CW (2014) {Planetesimal-driven
  migration as an explanation for observations of high levels of warm,
  exozodiacal dust}. \mnras 441:2380--2391, \doi{10.1093/mnras/stu721},
  \eprint{1404.2606}

\bibitem[{{Chen} et~al(2006){Chen}, {Sargent}, {Bohac}, {Kim}, {Leibensperger},
  {Jura}, {Najita}, {Forrest}, {Watson}, {Sloan}, and {Keller}}]{che06}
{Chen} CH, {Sargent} BA, {Bohac} C, {Kim} KH, {Leibensperger} E, {Jura} M,
  {Najita} J, {Forrest} WJ, {Watson} DM, {Sloan} GC, {Keller} LD (2006)
  {Spitzer IRS Spectroscopy of IRAS-discovered Debris Disks}. \apjs
  166:351--377, \doi{10.1086/505751}, \eprint{arXiv:astro-ph/0605277}

\bibitem[{{Defr{\`e}re} et~al(2010){Defr{\`e}re}, {Absil}, {den Hartog},
  {Hanot}, and {Stark}}]{def10}
{Defr{\`e}re} D, {Absil} O, {den Hartog} R, {Hanot} C, {Stark} C (2010)
  {Nulling interferometry: impact of exozodiacal clouds on the performance of
  future life-finding space missions}. \aap 509:A9,
  \doi{10.1051/0004-6361/200912973}, \eprint{0910.3486}

\bibitem[{{Defr{\`e}re} et~al(2011){Defr{\`e}re}, {Absil}, {Augereau}, {di
  Folco}, {Berger}, {Coud{\'e} du Foresto}, {Kervella}, {Le Bouquin},
  {Lebreton}, {Millan-Gabet}, {Monnier}, {Olofsson}, and {Traub}}]{def11}
{Defr{\`e}re} D, {Absil} O, {Augereau} JC, {di Folco} E, {Berger} JP,
  {Coud{\'e} du Foresto} V, {Kervella} P, {Le Bouquin} JB, {Lebreton} J,
  {Millan-Gabet} R, {Monnier} JD, {Olofsson} J, {Traub} W (2011) {Hot
  exozodiacal dust resolved around Vega with IOTA/IONIC}. \aap 534:A5,
  \doi{10.1051/0004-6361/201117017}, \eprint{1108.3698}

\bibitem[{{Defr{\`e}re} et~al(2012){Defr{\`e}re}, {Lebreton}, {Le Bouquin},
  {Lagrange}, {Absil}, {Augereau}, {Berger}, {di Folco}, {Ertel}, {Kluska},
  {Montagnier}, {Millan-Gabet}, {Traub}, and {Zins}}]{def12}
{Defr{\`e}re} D, {Lebreton} J, {Le Bouquin} JB, {Lagrange} AM, {Absil} O,
  {Augereau} JC, {Berger} JP, {di Folco} E, {Ertel} S, {Kluska} J, {Montagnier}
  G, {Millan-Gabet} R, {Traub} W, {Zins} G (2012) {Hot circumstellar material
  resolved around <ASTROBJ>{$\beta$} Pic</ASTROBJ> with VLTI/PIONIER}. \aap
  546:L9, \doi{10.1051/0004-6361/201220287}, \eprint{1210.1914}

\bibitem[{{Defr{\`e}re} et~al(2018){Defr{\`e}re}, {Absil}, {Berger}, {Boulet},
  {Danchi}, {Ertel}, {Gallenne}, {H{\'e}nault}, {Hinz}, {Huby}, {Ireland},
  {Kraus}, {Labadie}, {Le Bouquin}, {Martin}, {Matter}, {M{\'e}rand},
  {Mennesson}, {Minardi}, {Monnier}, {Norris}, {Orban de Xivry}, {Pedretti},
  {Pott}, {Reggiani}, {Serabyn}, {Surdej}, {Tristram}, and {Woillez}}]{def18}
{Defr{\`e}re} D, {Absil} O, {Berger} JP, {Boulet} T, {Danchi} WC, {Ertel} S,
  {Gallenne} A, {H{\'e}nault} F, {Hinz} P, {Huby} E, {Ireland} M, {Kraus} S,
  {Labadie} L, {Le Bouquin} JB, {Martin} G, {Matter} A, {M{\'e}rand} A,
  {Mennesson} B, {Minardi} S, {Monnier} J, {Norris} B, {Orban de Xivry} G,
  {Pedretti} E, {Pott} JU, {Reggiani} M, {Serabyn} E, {Surdej} J, {Tristram}
  KRW, {Woillez} J (2018) {The path towards high-contrast imaging with the
  VLTI: the Hi-5 project}. ArXiv e-prints \eprint{1801.04148}

\bibitem[{{Eiroa} et~al(2013){Eiroa}, {Marshall}, {Mora}, {Montesinos},
  {Absil}, {Augereau}, {Bayo}, {Bryden}, {Danchi}, {del Burgo}, {Ertel},
  {Fridlund}, {Heras}, {Krivov}, {Launhardt}, {Liseau}, {L{\"o}hne},
  {Maldonado}, {Pilbratt}, {Roberge}, {Rodmann}, {Sanz-Forcada}, {Solano},
  {Stapelfeldt}, {Th{\'e}bault}, {Wolf}, {Ardila}, {Ar{\'e}valo}, {Beichmann},
  {Faramaz}, {Gonz{\'a}lez-Garc{\'{\i}}a}, {Guti{\'e}rrez}, {Lebreton},
  {Mart{\'{\i}}nez-Arn{\'a}iz}, {Meeus}, {Montes}, {Olofsson}, {Su}, {White},
  {Barrado}, {Fukagawa}, {Gr{\"u}n}, {Kamp}, {Lorente}, {Morbidelli},
  {M{\"u}ller}, {Mutschke}, {Nakagawa}, {Ribas}, and {Walker}}]{eir13}
{Eiroa} C, {Marshall} JP, {Mora} A, {Montesinos} B, {Absil} O, {Augereau} JC,
  {Bayo} A, {Bryden} G, {Danchi} W, {del Burgo} C, {Ertel} S, {Fridlund} M,
  {Heras} AM, {Krivov} AV, {Launhardt} R, {Liseau} R, {L{\"o}hne} T,
  {Maldonado} J, {Pilbratt} GL, {Roberge} A, {Rodmann} J, {Sanz-Forcada} J,
  {Solano} E, {Stapelfeldt} K, {Th{\'e}bault} P, {Wolf} S, {Ardila} D,
  {Ar{\'e}valo} M, {Beichmann} C, {Faramaz} V, {Gonz{\'a}lez-Garc{\'{\i}}a} BM,
  {Guti{\'e}rrez} R, {Lebreton} J, {Mart{\'{\i}}nez-Arn{\'a}iz} R, {Meeus} G,
  {Montes} D, {Olofsson} G, {Su} KYL, {White} GJ, {Barrado} D, {Fukagawa} M,
  {Gr{\"u}n} E, {Kamp} I, {Lorente} R, {Morbidelli} A, {M{\"u}ller} S,
  {Mutschke} H, {Nakagawa} T, {Ribas} I, {Walker} H (2013) {DUst around NEarby
  Stars. The survey observational results}. \aap 555:A11,
  \doi{10.1051/0004-6361/201321050}, \eprint{1305.0155}

\bibitem[{{Ertel} et~al(2014){Ertel}, {Absil}, {Defr{\`e}re}, {Le Bouquin},
  {Augereau}, {Marion}, {Blind}, {Bonsor}, {Bryden}, {Lebreton}, and
  {Milli}}]{ert14b}
{Ertel} S, {Absil} O, {Defr{\`e}re} D, {Le Bouquin} JB, {Augereau} JC, {Marion}
  L, {Blind} N, {Bonsor} A, {Bryden} G, {Lebreton} J, {Milli} J (2014) {A
  near-infrared interferometric survey of debris-disk stars. IV. An unbiased
  sample of 92 southern stars observed in H band with VLTI/PIONIER}. \aap
  570:A128, \doi{10.1051/0004-6361/201424438}, \eprint{1409.6143}

\bibitem[{{Ertel} et~al(2016){Ertel}, {Defr{\`e}re}, {Absil}, {Le Bouquin},
  {Augereau}, {Berger}, {Blind}, {Bonsor}, {Lagrange}, {Lebreton}, {Marion},
  {Milli}, and {Olofsson}}]{ert16}
{Ertel} S, {Defr{\`e}re} D, {Absil} O, {Le Bouquin} JB, {Augereau} JC, {Berger}
  JP, {Blind} N, {Bonsor} A, {Lagrange} AM, {Lebreton} J, {Marion} L, {Milli}
  J, {Olofsson} J (2016) {A near-infrared interferometric survey of debris-disc
  stars. V. PIONIER search for variability}. \aap 595:A44,
  \doi{10.1051/0004-6361/201527721}, \eprint{1608.05731}

\bibitem[{{Ertel} et~al(2018{\natexlab{a}}){Ertel}, {Defr{\`e}re}, {Hinz},
  {Mennesson}, {Kennedy}, {Danchi}, {Gelino}, {Hill}, {Hoffmann}, {Rieke},
  {Shannon}, {Spalding}, {Stone}, {Vaz}, {Weinberger}, {Willems}, {Absil},
  {Arbo}, {Bailey}, {Beichman}, {Bryden}, {Downey}, {Durney}, {Esposito},
  {Gaspar}, {Grenz}, {Haniff}, {Leisenring}, {Marion}, {McMahon},
  {Millan-Gabet}, {Montoya}, {Morzinski}, {Pinna}, {Power}, {Puglisi},
  {Roberge}, {Serabyn}, {Skemer}, {Stapelfeldt}, {Su}, {Vaitheeswaran}, and
  {Wyatt}}]{ert18}
{Ertel} S, {Defr{\`e}re} D, {Hinz} P, {Mennesson} B, {Kennedy} GM, {Danchi} WC,
  {Gelino} C, {Hill} JM, {Hoffmann} WF, {Rieke} G, {Shannon} A, {Spalding} E,
  {Stone} JM, {Vaz} A, {Weinberger} AJ, {Willems} P, {Absil} O, {Arbo} P,
  {Bailey} VP, {Beichman} C, {Bryden} G, {Downey} EC, {Durney} O, {Esposito} S,
  {Gaspar} A, {Grenz} P, {Haniff} CA, {Leisenring} JM, {Marion} L, {McMahon}
  TJ, {Millan-Gabet} R, {Montoya} M, {Morzinski} KM, {Pinna} E, {Power} J,
  {Puglisi} A, {Roberge} A, {Serabyn} E, {Skemer} AJ, {Stapelfeldt} K, {Su}
  KYL, {Vaitheeswaran} V, {Wyatt} MC (2018{\natexlab{a}}) {The HOSTS Survey --
  Exozodiacal Dust Measurements for 30 Stars}. \aj 155:194,
  \doi{10.3847/1538-3881/aab717}, \eprint{1803.11265}

\bibitem[{{Ertel} et~al(2018{\natexlab{b}}){Ertel}, {Kennedy}, {Defr{\`e}re},
  and {et al.}}]{ert18b}
{Ertel} S, {Kennedy} GM, {Defr{\`e}re} D, {et al} (2018{\natexlab{b}}) {The
  HOSTS Survey for Exo-zodiacal Dust: Preliminary results and future
  prospects}, proceedings SPIE, in press

\bibitem[{{Faramaz} et~al(2017){Faramaz}, {Ertel}, {Booth}, {Cuadra}, and
  {Simmonds}}]{far17}
{Faramaz} V, {Ertel} S, {Booth} M, {Cuadra} J, {Simmonds} C (2017) {Inner
  mean-motion resonances with eccentric planets: a possible origin for
  exozodiacal dust clouds}. \mnras 465:2352--2365, \doi{10.1093/mnras/stw2846},
  \eprint{1611.02196}

\bibitem[{{Hahn} et~al(2002){Hahn}, {Zook}, {Cooper}, and {Sunkara}}]{hah02}
{Hahn} JM, {Zook} HA, {Cooper} B, {Sunkara} B (2002) {Clementine Observations
  of the Zodiacal Light and the Dust Content of the Inner Solar System}.
  \icarus 158:360--378, \doi{10.1006/icar.2002.6881}, \eprint{astro-ph/0204111}

\bibitem[{{Kennedy} and {Wyatt}(2013)}]{ken13}
{Kennedy} GM, {Wyatt} MC (2013) {The bright end of the exo-Zodi luminosity
  function: disc evolution and implications for exo-Earth detectability}.
  \mnras 433:2334--2356, \doi{10.1093/mnras/stt900}

\bibitem[{{Kimura} and {Mann}(1998)}]{kim98}
{Kimura} H, {Mann} I (1998) {Brightness of the solar F-corona}. Earth, Planets,
  and Space 50:493--499

\bibitem[{{Kirchschlager} et~al(2017){Kirchschlager}, {Wolf}, {Krivov},
  {Mutschke}, and {Brunngr{\"a}ber}}]{kir17}
{Kirchschlager} F, {Wolf} S, {Krivov} AV, {Mutschke} H, {Brunngr{\"a}ber} R
  (2017) {Constraints on the structure of hot exozodiacal dust belts}. \mnras
  467:1614--1626, \doi{10.1093/mnras/stx202}, \eprint{1701.07271}

\bibitem[{{Kirchschlager} et~al(2018){Kirchschlager}, {Wolf},
  {Brunngr{\"a}ber}, {Matter}, {Krivov}, and {Labdon}}]{kir18}
{Kirchschlager} F, {Wolf} S, {Brunngr{\"a}ber} R, {Matter} A, {Krivov} AV,
  {Labdon} A (2018) {Modelling of mid-infrared interferometric signature of hot
  exozodiacal dust emission}. \mnras 473:2633--2638,
  \doi{10.1093/mnras/stx2515}, \eprint{1709.08514}

\bibitem[{{Kral} et~al(2017){Kral}, {Krivov}, {Defrere}, {van Lieshout},
  {Bonsor}, {Augereau}, {Thebault}, {Absil}, and {Ertel}}]{kra17}
{Kral} Q, {Krivov} AV, {Defrere} D, {van Lieshout} R, {Bonsor} A, {Augereau}
  JC, {Thebault} P, {Absil} O, {Ertel} S (2017) {Exozodiacal clouds: Hot and
  warm dust around main sequence stars}. ArXiv e-prints \eprint{1703.02540}

\bibitem[{{Lebreton} et~al(2013){Lebreton}, {van Lieshout}, {Augereau},
  {Absil}, {Mennesson}, {Kama}, {Dominik}, {Bonsor}, {Vandeportal}, {Beust},
  {Defr{\`e}re}, {Ertel}, {Faramaz}, {Hinz}, {Kral}, {Lagrange}, {Liu}, and
  {Th{\'e}bault}}]{leb13}
{Lebreton} J, {van Lieshout} R, {Augereau} JC, {Absil} O, {Mennesson} B, {Kama}
  M, {Dominik} C, {Bonsor} A, {Vandeportal} J, {Beust} H, {Defr{\`e}re} D,
  {Ertel} S, {Faramaz} V, {Hinz} P, {Kral} Q, {Lagrange} AM, {Liu} W,
  {Th{\'e}bault} P (2013) {An interferometric study of the Fomalhaut inner
  debris disk. III. Detailed models of the exozodiacal disk and its origin}.
  \aap 555:A146, \doi{10.1051/0004-6361/201321415}, \eprint{1306.0956}

\bibitem[{{Lisse} et~al(2007){Lisse}, {Beichman}, {Bryden}, and
  {Wyatt}}]{lis07}
{Lisse} CM, {Beichman} CA, {Bryden} G, {Wyatt} MC (2007) {On the Nature of the
  Dust in the Debris Disk around HD 69830}. \apj 658:584--592,
  \doi{10.1086/511001}, \eprint{arXiv:astro-ph/0611452}

\bibitem[{{Lisse} et~al(2008){Lisse}, {Chen}, {Wyatt}, and {Morlok}}]{lis08}
{Lisse} CM, {Chen} CH, {Wyatt} MC, {Morlok} A (2008) {Circumstellar Dust
  Created by Terrestrial Planet Formation in HD 113766}. \apj 673:1106--1122,
  \doi{10.1086/523626}, \eprint{0710.0839}

\bibitem[{{Marion} et~al(2014){Marion}, {Absil}, {Ertel}, {Le Bouquin},
  {Augereau}, {Blind}, {Defr{\`e}re}, {Lebreton}, and {Milli}}]{mar14}
{Marion} L, {Absil} O, {Ertel} S, {Le Bouquin} JB, {Augereau} JC, {Blind} N,
  {Defr{\`e}re} D, {Lebreton} J, {Milli} J (2014) {Searching for faint
  companions with VLTI/PIONIER. II. 92 main sequence stars from the Exozodi
  survey}. \aap 570:A127, \doi{10.1051/0004-6361/201424780}, \eprint{1409.6105}

\bibitem[{{Marshall} et~al(2016){Marshall}, {Cotton}, {Bott}, {Ertel},
  {Kennedy}, {Wyatt}, {del Burgo}, {Absil}, {Bailey}, and
  {Kedziora-Chudczer}}]{mar16}
{Marshall} JP, {Cotton} DV, {Bott} K, {Ertel} S, {Kennedy} GM, {Wyatt} MC, {del
  Burgo} C, {Absil} O, {Bailey} J, {Kedziora-Chudczer} L (2016) {Polarization
  Measurements of Hot Dust Stars and the Local Interstellar Medium}. \apj
  825:124, \doi{10.3847/0004-637X/825/2/124}, \eprint{1604.08286}

\bibitem[{{Mennesson} et~al(2014){Mennesson}, {Millan-Gabet}, {Serabyn},
  {Colavita}, {Absil}, {Bryden}, {Wyatt}, {Danchi}, {Defr{\`e}re}, {Dor{\'e}},
  {Hinz}, {Kuchner}, {Ragland}, {Scott}, {Stapelfeldt}, {Traub}, and
  {Woillez}}]{men14}
{Mennesson} B, {Millan-Gabet} R, {Serabyn} E, {Colavita} MM, {Absil} O,
  {Bryden} G, {Wyatt} M, {Danchi} W, {Defr{\`e}re} D, {Dor{\'e}} O, {Hinz} P,
  {Kuchner} M, {Ragland} S, {Scott} N, {Stapelfeldt} K, {Traub} W, {Woillez} J
  (2014) {Constraining the Exozodiacal Luminosity Function of Main-sequence
  Stars: Complete Results from the Keck Nuller Mid-infrared Surveys}. \apj
  797:119, \doi{10.1088/0004-637X/797/2/119}

\bibitem[{{Millan-Gabet} et~al(2011){Millan-Gabet}, {Serabyn}, {Mennesson},
  {Stark}, {Ragland}, {Hrynevych}, {Woillez}, {Stapelfeldt}, {Bryden},
  {Colavita}, and {Booth}}]{mil11}
{Millan-Gabet} R, {Serabyn} E, {Mennesson} B, {Stark} CC, {Ragland} S,
  {Hrynevych} M, {Woillez} J, {Stapelfeldt} K, {Bryden} G, {Colavita} MM,
  {Booth} AJ (2011) {Exozodiacal Dust Levels for Nearby Main-sequence Stars: A
  Survey with the Keck Interferometer Nuller}. \apj 734:67,
  \doi{10.1088/0004-637X/734/1/67}, \eprint{1104.1382}

\bibitem[{{Nu\~nez} et~al(2017){Nu\~nez}, {Scott}, {Mennesson}, {Absil},
  {Augereau}, {Bryden}, {ten Brummelaar}, {Ertel}, {Coude du Foresto},
  {Ridgway}, {Sturmann}, {Sturmann}, {Turner}, and {Turner}}]{nun17}
{Nu\~nez} PD, {Scott} NJ, {Mennesson} B, {Absil} O, {Augereau} JC, {Bryden} G,
  {ten Brummelaar} T, {Ertel} S, {Coude du Foresto} V, {Ridgway} ST, {Sturmann}
  J, {Sturmann} L, {Turner} NJ, {Turner} NH (2017) {A near-infrared
  interferometric survey of debris-disk stars. VI. Extending the exozodiacal
  light survey with CHARA/JouFLU}. ArXiv e-prints \eprint{1709.01655}

\bibitem[{{Olofsson} et~al(2012){Olofsson}, {Juh{\'a}sz}, {Henning},
  {Mutschke}, {Tamanai}, {Mo{\'o}r}, and {{\'A}brah{\'a}m}}]{olo12}
{Olofsson} J, {Juh{\'a}sz} A, {Henning} T, {Mutschke} H, {Tamanai} A,
  {Mo{\'o}r} A, {{\'A}brah{\'a}m} P (2012) {Transient dust in warm debris
  disks. Detection of Fe-rich olivine grains}. \aap 542:A90,
  \doi{10.1051/0004-6361/201118735}, \eprint{1204.2374}

\bibitem[{{Rieke} et~al(2016){Rieke}, {G{\'a}sp{\'a}r}, and
  {Ballering}}]{rie16}
{Rieke} GH, {G{\'a}sp{\'a}r} A, {Ballering} NP (2016) {Magnetic Grain Trapping
  and the Hot Excesses around Early-type Stars}. \apj 816:50,
  \doi{10.3847/0004-637X/816/2/50}, \eprint{1511.04998}

\bibitem[{{Roberge} et~al(2012){Roberge}, {Chen}, {Millan-Gabet}, {Weinberger},
  {Hinz}, {Stapelfeldt}, {Absil}, {Kuchner}, and {Bryden}}]{rob12}
{Roberge} A, {Chen} CH, {Millan-Gabet} R, {Weinberger} AJ, {Hinz} PM,
  {Stapelfeldt} KR, {Absil} O, {Kuchner} MJ, {Bryden} G (2012) {The Exozodiacal
  Dust Problem for Direct Observations of Exo-Earths}. \pasp 124:799--808,
  \doi{10.1086/667218}, \eprint{1204.0025}

\bibitem[{{Stark} et~al(2015){Stark}, {Roberge}, {Mandell}, {Clampin},
  {Domagal-Goldman}, {McElwain}, and {Stapelfeldt}}]{sta15}
{Stark} CC, {Roberge} A, {Mandell} A, {Clampin} M, {Domagal-Goldman} SD,
  {McElwain} MW, {Stapelfeldt} KR (2015) {Lower Limits on Aperture Size for an
  ExoEarth Detecting Coronagraphic Mission}. \apj 808:149,
  \doi{10.1088/0004-637X/808/2/149}, \eprint{1506.01723}

\bibitem[{{Stock} et~al(2010){Stock}, {Su}, {Liu}, {Hinz}, {Rieke}, {Marengo},
  {Stapelfeldt}, {Hines}, and {Trilling}}]{sto10}
{Stock} ND, {Su} KYL, {Liu} W, {Hinz} PM, {Rieke} GH, {Marengo} M,
  {Stapelfeldt} KR, {Hines} DC, {Trilling} DE (2010) {The Structure of the
  {$\beta$} Leonis Debris Disk}. \apj 724:1238--1255,
  \doi{10.1088/0004-637X/724/2/1238}, \eprint{1010.0003}

\end{thebibliography}

\end{document}